# Distributed Circuit Model for Predicting the Quality Factor of Magnetic Polariton Resonance

Hangjie Li[1], Junming Zhao[1,2*]

[1] *School of Energy Science and Engineering, Harbin Institute of Technology, Harbin 150001, China*

[2] *Key Laboratory of Aerospace Thermophysics, Ministry of Industry and Information Technology of China, Harbin 150001, China*

## Abstract

Existing RLC circuit model leads to inaccurate predictions of the quality factor (Q-factor) of magnetic polariton (MP) resonances under imperfect absorption conditions due to the omission of radiation loss. Moreover, the lumped-parameter nature of RLC models also limits their applicability for predicting higher-order MP modes. In this letter, we propose a distributed circuit model (DCM) for predicting the Q-factor of MP resonances, which overcomes these limitations. By introducing a radiation resistance to characterize radiation loss and establishing a mapping between distributed and lumped parameters, we derive a unified analytical expression for the Q-factor of arbitrary-order MPs. Validation via rigorous coupled-wave analysis (RCWA) demonstrates that this model can accurately predict the Q-factors of MPs of all orders in various metal–insulator–metal (MIM) structures. This work provides a simple yet effective tool for designing metamaterial emitters/absorbers and advances the understanding of MP loss mechanisms.

* Corresponding author.

*Email address*: jmzhao@hit.edu.cn

Metamaterials with subwavelength features enable spectral control of thermal radiation beyond conventional broadband emitters.[1] Magnetic polaritons (MPs), as one of the fundamental resonant modes, provide highly selective resonances and have been widely used in solar energy harvesting,[2-4] thermophotovoltaic systems,[5-7] radiative cooling,[8-10] and thermal camouflage.[11-13] Equivalent circuit models have proven to be an effective tool for predicting the resonance conditions of MPs in micro/nanostructures,[14-16] in which the resonance frequency of the MP is identified by minimizing the magnitude of the total impedance of the equivalent circuit, i.e., $\omega_{\mathrm{R}} = \arg\min\left|\tilde{Z}(\omega)\right|$.[17-19] In metal–insulator–metal (MIM) structures, the resonances are also interpreted as gap plasmon (GP) as well.[20-25]

Previous studies primarily focused on the prediction of the MP resonance frequency by using the lumped circuit model (LCM), whereas the prediction of quality factor (Q-factor) and full width at half maximum (FWHM) of MP resonance remain largely unexplored.[26-29] Sakurai et al.[26] initially proposed an RLC equivalent circuit model for describing MPs in MIM infrared absorbers, which enables simultaneous prediction of the resonance frequency, Q-factor, and FWHM. However, this model is accurate only for the resonance near perfect absorption. Substantial deviations in the predicted Q-factor and FWHM are observed when significant transmission/reflection occurs at resonance. This discrepancy arises because the model neglects radiation losses caused by scattering, in which the resistance $R$ is only the Ohmic loss, calculated from the kinetic impedance $Z_{\mathrm{k}}$ associated with the kinetic energy of electrons inside the metal. It will be analyzed in detail in the subsequent section of this letter. Moreover, extending the LCM to higher-order MP modes is difficult and cumbersome because a separate equivalent circuit must be constructed for each mode.[29-32] Most recently, a distributed circuit model (DCM) was proposed to predict the resonance frequencies of multiple MP modes within a unified circuit framework.[17] Even though, predicting the Q-factor and FWHM of MP resonance for multiple MP modes remains very challenging due to the lack of radiation loss model.

In this letter, we propose a DCM to predict the Q-factors of arbitrary-order MP resonances. To account for radiative losses, a radiation resistance $R_{\mathrm{rad}}$, is incorporated into the DCM. By establishing a mapping between the distributed circuit parameters and the corresponding lumped circuit parameters, a unified analytical expression for the Q-factors of arbitrary-order MPs is derived. The theoretical predictions are further compared with the results obtained from rigorous coupled-wave analysis (RCWA) to validate the accuracy of the model.

We first consider the case of perfect MP absorption. For the substrate-supported horizontal MIM structure shown in **Fig. 1**(a), the geometry is characterized by the period Λ, metal strip thickness $w$, slit length $h$, and slit width $b$. To realize perfect absorption, the structural parameters are chosen as follows: $\Lambda$ = 1.2 μm, $w$ = 0.2 μm, $h$ = 0.8 μm, and $b$ = 0.02 μm. Ag is employed for both the metallic strip and substrate, whereas ZnSe is used as the dielectric spacer. The optical response of Ag is described by the Drude model $\tilde{\varepsilon}(\omega)=\varepsilon_\infty-\omega_p{}^2/(\omega^2-\mathrm{i}\Gamma\omega)$ with the following parameters: plasma frequency $\omega_p$ = $1.39\times10^{16}$ rad/s, scattering rate $\Gamma$ = $2.70\times10^{13}$ rad/s, and a high-frequency constant $\varepsilon_\infty$ = 3.4.[15] ZnSe is treated as a nondispersive dielectric (refractive index $n$ = 2.4).[33] The structure is surrounded by free space, and only TM-polarized waves are considered. **Fig. 1**(b) presents the radiative characteristics under normal incidence calculated by RCWA. RCWA calculations employed 181 Fourier orders to ensure convergence. [34] Under this condition, the peak absorptance of the fundamental MP mode (MP1) reaches unity. In **Fig. 1**(b), the diamond denotes the Q-factor predicted by the RLC model proposed by Sakurai.[26] The RLC prediction agrees well with the numerical results computed from RCWA simulation. Because the FWHM is directly related to the resonance frequency $\omega_R$ and the Q-factor, only the predicted Q-factors are shown in the figure.

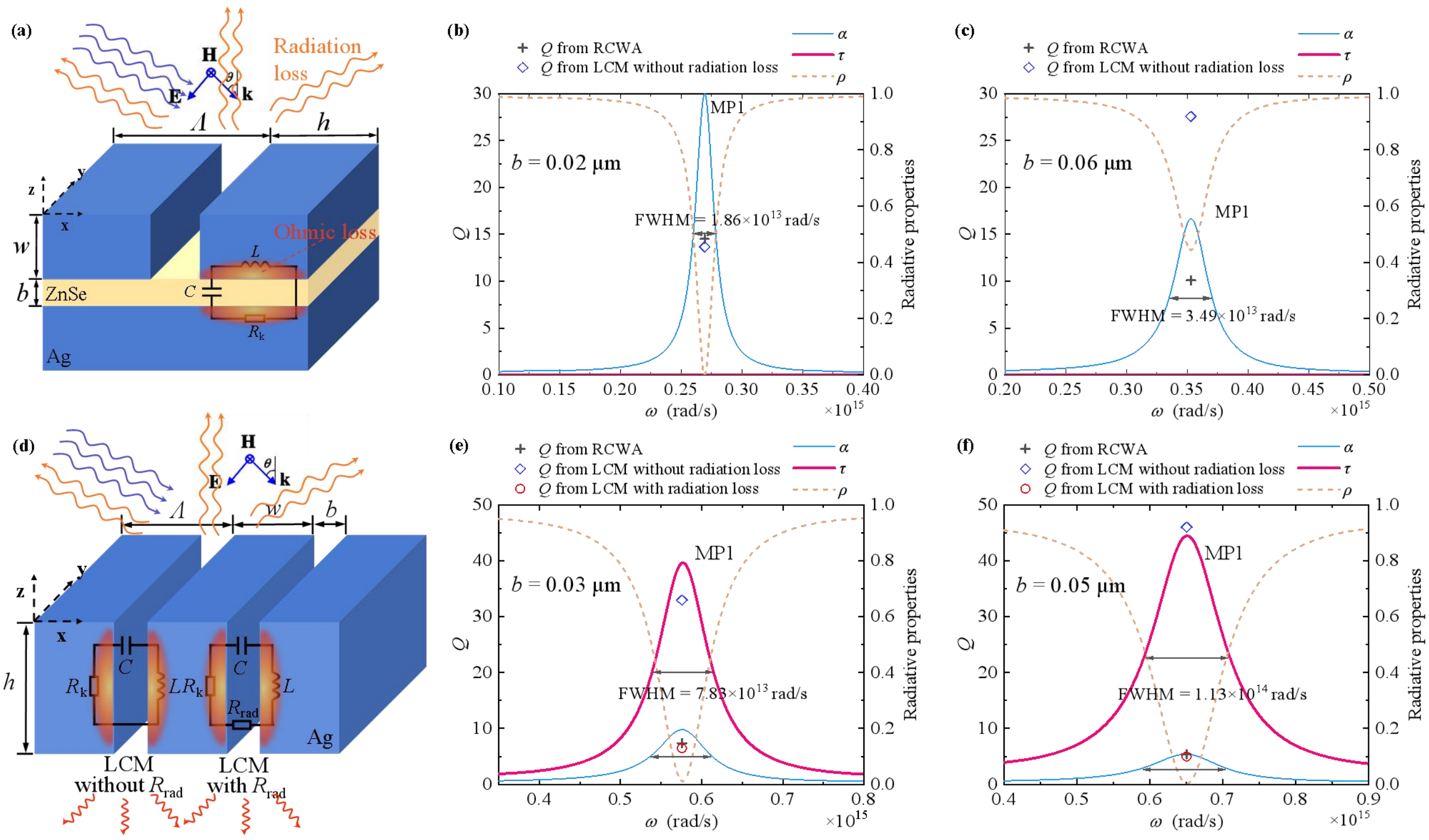

**Fig. 1** Q-factor prediction for MP1 using the RLC model. (a) Schematic of the substrate-supported horizontal MIM structure. (b, c) Q-factors for MP1 obtained from the RLC model without $R_{rad}$ (denoted by diamonds, ◇) and from RCWA (denoted by crosses, +) for slit widths of $b$=0.02 μm (b) and $b$=0.06μm (c). (d) Schematic of the substrate-free vertical MIM structure. (e, f) Q-factors obtained from the RLC model with $R_{rad}$ (denoted by circles, ○) and from RCWA (+) for slit widths of $b$=0.03μm (e) and $b$=0.05μm (f), respectively. In the spectral plots, the thin solid blue line, thick solid magenta line, and dashed orange line denote the absorptance $\alpha$, transmittance $\tau$, and reflectance $\rho$ under normal incidence, respectively.

Increasing only the slit width to $b$ = 0.06 μm reduces the MP1 peak absorptance well below unity and produces a clear discrepancy between the RLC model and RCWA, as shown in **Fig. 1**(c). The reason is that the resistance $R$ in their model is derived solely from the kinetic impedance $Z_k$ associated with the kinetic energy of electrons inside the metal, [26]

$$Z_k = R_k - i\omega L_k \tag{1}$$

Thus, the model contains only the Ohmic kinetic resistance $R_k$ and ignores radiative loss. This approximation is valid near perfect absorption but fails when radiative loss becomes comparable to Ohmic dissipation.

Significant radiative losses are inherent in the substrate-free vertical MIM structure shown in **Fig. 1**(d), where the slit is filled with air and enclosed by Ag on both sides. Consequently, the original RLC model, which neglects radiative losses, yields substantial prediction errors. The key to resolving this issue lies in incorporating radiative losses into the RLC model. To this end, we introduce a radiation resistance $R_{rad}$, to characterize radiative losses and connect it in series with the kinetic resistance $R_k$ in the original RLC model, as illustrated for the right slit in **Fig. 1**(d). For a series RLC circuit operating at resonance, the Q-factor is conventionally expressed as $Q = \frac{\omega_R L}{R}$. By incorporating radiative losses into the circuit model, the quality factor can be rewritten as:

$$Q = \frac{\omega_R L}{R_{rad} + R_k} = \frac{1}{R_{rad} + R_k}\sqrt{\frac{L}{C}} \tag{2}$$

where $\omega_R$ is the resonance angular frequency of the series RLC circuit $\omega_R = 1/\sqrt{LC}$. The remaining challenge is to determine the radiation resistance $R_{rad}$. In a series RLC circuit, the dissipated power is proportional to the corresponding resistance $P = I^2 R$. At resonance, the peak absorptance $\alpha_R$ can be interpreted as the fraction of the total dissipated power converted into Ohmic losses. Therefore, we can reasonably assume that the ratio between the radiation resistance $R_{rad}$ and the kinetic resistance $R_k$ is given by

$$\frac{R_{rad}}{R_k} = \frac{1-\alpha_R}{\alpha_R} \tag{3}$$

where $\alpha_R$ is the peak absorptance of the MP resonance. With the kinetic resistance $R_k$ already available and the peak absorptance $\alpha_R$ extracted from the spectral response, the radiation resistance $R_{rad}$ can be readily obtained from Eq. (3).

The RLC model incorporating radiation resistance $R_{rad}$ significantly improves the prediction accuracy. It reduces the relative errors from 348.29% to 11.76% and from 752.13% to 7.65% for $b$ =

0.03 and 0.05 μm, respectively, as shown in **Fig. 1** (e, f). As $b$ increases, $\alpha_R$ decreases, $R_{rad}$ / $R_k$ increases, and the error of the conventional RLC model becomes larger. This result further validates the rationality of the proposed radiation resistance $R_{rad}$. To date, studies on MP Q-factors based on equivalent circuit models have been limited to the prediction of the MP1 using LCMs.

To predict the Q-factors of higher-order MP resonances, we extend the DCM in Ref. 17 by incorporating radiative losses, as illustrated for the left slit in **Fig. 2**(a). The distributed kinetic resistance $R_{k0}$, distributed inductance $L_0$, and distributed capacitance $C_0$ are taken from Ref. 17. Here, we further introduce a distributed radiation resistance $R_{rad0}$ to account for radiative losses. When the $m$th-order magnetic polariton resonance (MP$_m$) is excited, $m$ current loops with identical magnitudes are formed around the slit,[35] as illustrated in the inset on the right side of **Fig. 2**(b). Each loop can be represented by an equivalent RLC circuit with the same resonance frequency.[30] The key idea is to establish a mapping between the distributed circuit parameters and the corresponding lumped circuit parameters, so that the Q-factors of arbitrary-order MP resonances can be predicted within a unified framework using a single equivalent lumped circuit. For the substrate-free vertical MIM structure shown in **Fig. 2**(a), the resonance condition for the $m$th-order MP mode can be expressed as[17]

$$\begin{aligned}\omega_{\mathrm{MP}m} &= \frac{m\pi}{h\sqrt{L_0(\omega_{\mathrm{MP}m})C_0(\omega_{\mathrm{MP}m})}} \\ &= \frac{1}{\sqrt{\dfrac{hL_0(\omega_{\mathrm{MP}m})}{m\pi}\dfrac{hC_0(\omega_{\mathrm{MP}m})}{m\pi}}}\ , \quad m = 1, 2, \ldots\end{aligned} \tag{4}$$

where $m$ is the resonance order of the MP mode. By drawing an analogy with the resonance frequency of a series RLC lumped circuit $\omega_R = 1/\sqrt{LC}$ , the following expression is obtained:

$$L(\omega_{\mathrm{MP}m}) = \frac{hL_0(\omega_{\mathrm{MP}m})}{m\pi} \tag{5}$$

$$C(\omega_{\mathrm{MP}m}) = \frac{hC_0(\omega_{\mathrm{MP}m})}{m\pi} \tag{6}$$

Accordingly, the expression for the kinetic resistance $R_k$ is obtained as:

$$R_k(\omega_{\mathrm{MP}m}) = \frac{hR_{k0}(\omega_{\mathrm{MP}m})}{m\pi} \tag{7}$$

Assuming a uniform distribution of loss energy among all current loops, the kinetic resistance $R_k$ and radiation resistance $R_{rad}$ are identical for each lumped circuit. In this case, $R_{rad}$ and $R_k$ still satisfy Eq. (3). Therefore, $R_{rad}$ can be written in a form similar to that of $R_k$:

$$R_{\mathrm{rad}}(\omega_{\mathrm{MP}m}) = \frac{hR_{\mathrm{rad0}}(\omega_{\mathrm{MP}m})}{m\pi} \tag{8}$$

where $R_{\mathrm{rad0}} = c_1 \frac{(1-\alpha_{\mathrm{R}})/m}{\alpha_{\mathrm{R}}/m} R_{\mathrm{k0}}$ , and $c_1$ is an energy-distribution correction factor. For an approximately uniform distribution of loss among all current loops, $c_1 = 1$. Thus, the mapping between the lumped and distributed circuit parameters is fully established. Substituting these relations into Eq. (2) yields the Q-factor of arbitrary-order MP resonances, which can be further simplified as:

$$Q_{\mathrm{MP}m} = \frac{\omega_{\mathrm{MP}m} L(\omega_{\mathrm{MP}m})}{R_{\mathrm{rad}}(\omega_{\mathrm{MP}m}) + R_{\mathrm{k}}(\omega_{\mathrm{MP}m})} = \frac{\omega_{\mathrm{MP}m} L_0(\omega_{\mathrm{MP}m})}{c_1 R_{\mathrm{rad0}}(\omega_{\mathrm{MP}m}) + R_{\mathrm{k0}}(\omega_{\mathrm{MP}m})} \tag{9}$$

where $\omega_{\mathrm{MP}m}$ represents the resonance frequency of the $m$th-order MP mode, which can be determined from Eq. (4). According to Eq. (9), the Q-factors of arbitrary-order MP resonances can be directly determined from the distributed circuit parameters alone.

Next, we validate the accuracy of the proposed model. Since the resonance frequency of the SPP remains unchanged when the period and incident angle are fixed,[36] a relatively small period Λ and slit width $b$, together with a sufficiently large slit length $h$, were chosen to separate the higher-order MP resonances from the SPP resonance frequency, thus minimizing the influence of SPP–MP coupling. The Q-factors obtained from the absorption spectra (+) and the transmission spectra (□) are identical for all MP resonances, as shown in **Fig. 2**(b). Since $\alpha + \tau + \rho = 1$, the Q-factors extracted from the reflection spectra are also identical. This is because the absorption enhancement, transmission enhancement, and reflection dip all originate from the same MP resonance. In **Fig. 2**(b), the circles correspond to the predictions of the DCM incorporating radiative losses, with $c_1$=1. The incorporation of radiation loss reduces the mean relative error (MRE) from 504.85% to 9.14%. A relatively larger discrepancy is observed only for MP5 because of its proximity to the SPP mode.

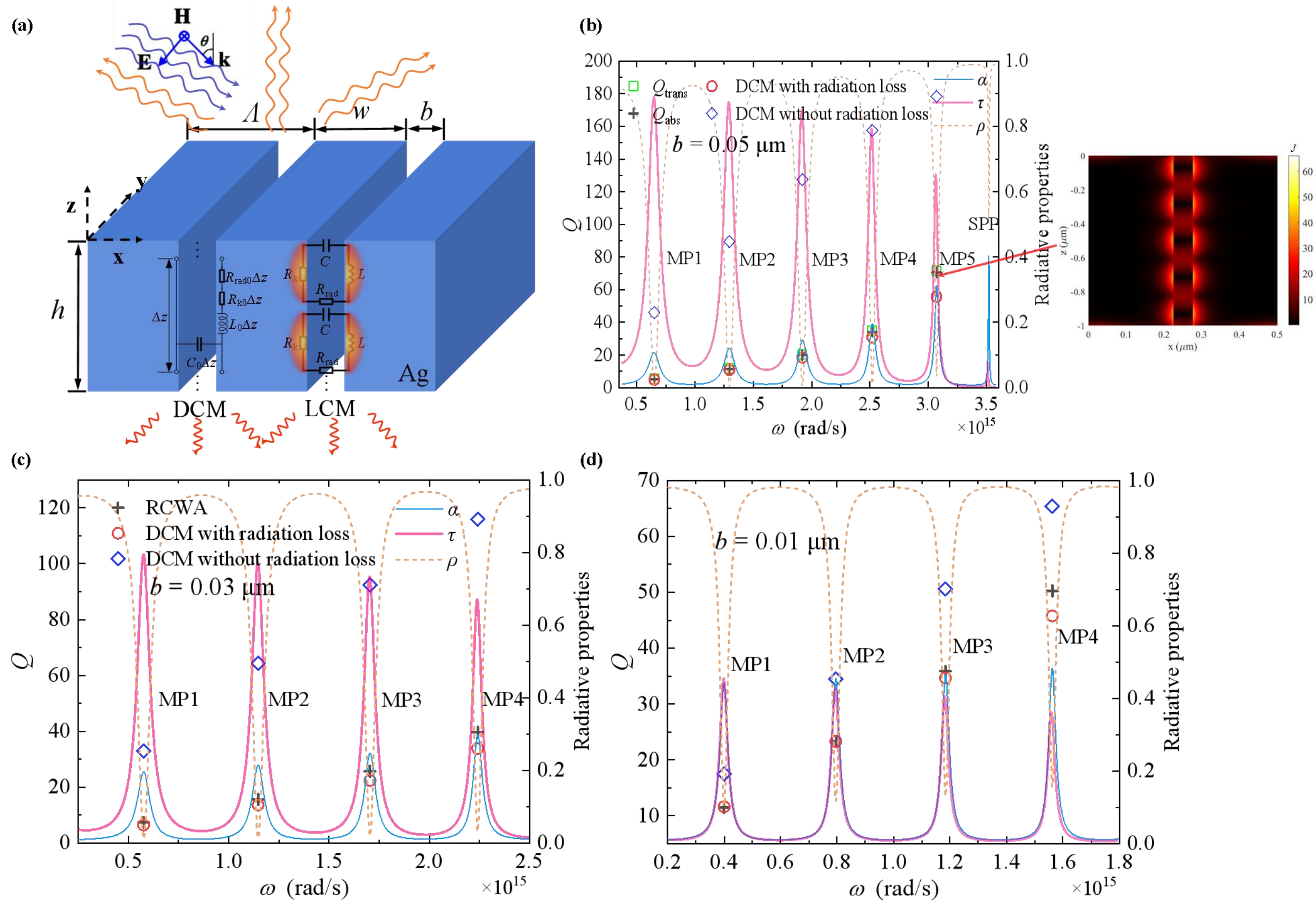

**Fig. 2** DCM for predicting the Q-factors of multi-order MPs. (a) Schematic of the vertical MIM structure without a substrate. (b) Q-factors obtained from the absorptance spectra (denoted by crosses, +) and transmittance spectra (denoted by squares, □), as well as predicted by the DCM with radiation loss (denoted by circles, ○) and without radiation loss (denoted by diamonds, ◇) at $\Lambda$=0.5μm, $h$=1μm, and $b$=0.05μm with $c_1$=1. The inset shows the dimensionless current density distribution for the MP5 resonance. The dimensionless current density is defined as $J=\left|\mathrm{j}\omega\varepsilon_0\tilde{\varepsilon}\tilde{E}/\mathrm{j}\omega\varepsilon_0\tilde{E}_{\mathrm{inc}}\right|=\left|\tilde{\varepsilon}\tilde{E}/\tilde{E}_{\mathrm{inc}}\right|$. (c) Results at $b$=0.03μm with $c_1$=1; (d) Results at $b$=0.01μm with $c_1$=0.4.

Further validation is shown in **Fig. 2** (c) and (d). For $b$ = 0.03 μm, $c_1$ = 1 remains sufficient. For $b$ = 0.01 μm, $c_1$ = 0.4 is required. When the slit width $b$ becomes sufficiently small, the loss distribution deviates from the assumption of uniformity. In fact, for higher-order MP resonances, the total radiative loss arises from the superposition of radiation emitted by all current loops, and radiation from adjacent slits may also interact with each other. Under such circumstances, the proposed approach can be regarded as extracting an effective lumped circuit that incorporates these collective radiative effects, thereby enabling accurate prediction of the Q-factors.

We further apply the proposed model to substrate-supported horizontal MIM structures as shown in **Fig. 3**(a). In this structure, the resonance conditions of the MP modes are identical to those of the vertical MIM structure.[17] Therefore, Eq. (9) is still employed to predict the Q-factors. However, owing to the structural asymmetry introduced by the substrate, the loss distribution inevitably deviates from the assumption of uniformity and must be corrected by the coefficient $c_1$. The predicted Q-factors of the MP resonances under normal incidence for different geometric configurations are presented in **Fig. 3**. Since only odd-order MP modes can be excited in the horizontal MIM structure under normal incidence, only the prediction results for MP1, MP3, and MP5 are shown. The inset in **Fig. 3**(c) illustrates the dimensionless current density distribution within the structure when MP5 is excited. Similar to the vertical MIM structure, all current loops exhibit nearly identical magnitudes. Ge with $n$ = 4 is used as the spacer to shift higher-order MPs away from the SPP resonance.[33]

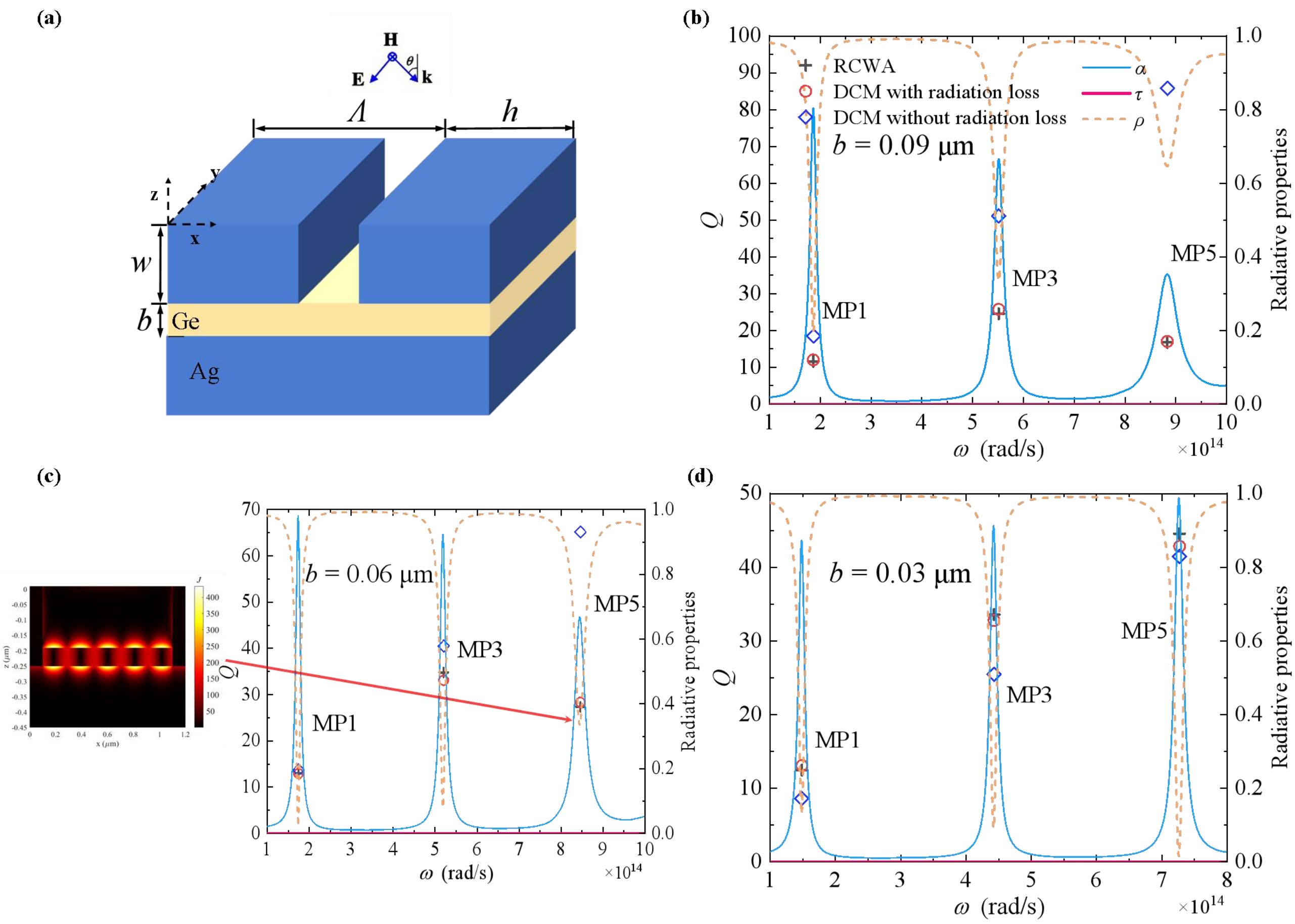

**Fig. 3** Validation of the DCM predictions for Q-factors of multi-order MPs in substrate-supported horizontal MIM structures. (a) Schematic of the substrate-supported horizontal MIM structure; (b) Results at Λ=1.2μm, $w$=0.2μm, $h$=1μm, and slit width $b$=0.09μm with $c_1$=2.2; (c) Results at $b$=0.06μm with $c_1$=2.6 (the inset shows the dimensionless current density distribution for the MP5 resonance); (d) Results at $b$=0.03μm with $c_1$ = −2.3. All results are obtained under normal incidence. The line styles and symbols are defined in (b) and used consistently in (c) and (d).

The results presented in **Fig. 3** clearly demonstrate that the incorporation of radiative loss is essential for accurately reproducing the Q-factor in horizontal MIM structures. With Λ = 1.2 μm, $w$ = 0.2 μm, and $h$ = 1 μm fixed, **Fig. 3**(b) presents the prediction results for $b$ = 0.09 μm, for which the correction coefficient $c_1$ is taken as 2.2. The MRE decreases from 192.77% to 3.25% when radiation loss is included. **Fig. 3**(c) shows the corresponding results for $b$ = 0.06 μm, where $c_1$ = 2.6. Under these conditions, the Q-factor does not increase monotonically with increasing MP resonance order. The proposed model captures this nonmonotonic behavior, whereas neglecting radiative losses leads to a monotonic increase of Q-factors with MP order. When the slit width is further reduced to $b$ = 0.03 μm, the prediction results are shown in **Fig. 3**(d). In this case, a negative correction coefficient, $c_1$ = −2.3, is required to accurately predict the Q-factors. The dependence of the Q-factor on the MP resonance order differs significantly from the previous cases and exhibits a monotonic increasing trend. Moreover, when radiative losses are neglected, the predicted Q-factors are smaller than those obtained from RCWA, indicating that the kinetic resistance $R_k$ is overestimated under this condition.

We further apply the proposed model to substrate-supported vertical MIM structures as shown in **Fig. 4**(a). When the *m*th-order magnetic polariton resonance (MP*m*) is excited, the magnitudes of the upper *m* - 1 loops are identical. [35] In contrast, the bottom loop shows significantly different characteristics due to the substrate, as depicted in the inset of **Fig. 4**(b). Its resonance condition is[17]

$$\begin{aligned}\omega_{\mathrm{MP}m} &= \frac{\xi_m}{h\sqrt{L_0(\omega_{\mathrm{MP}m})C_0(\omega_{\mathrm{MP}m})}} \\ &= \frac{1}{\sqrt{\dfrac{hL_0(\omega_{\mathrm{MP}m})}{\xi_m}\dfrac{hC_0(\omega_{\mathrm{MP}m})}{\xi_m}}}\end{aligned} \tag{9}$$

where, $\xi_m$ is the root of the transcendental equation ( $\cot x = kx$ ) in the $m$-th period ($m\pi - \pi$, $m\pi$), $k = b/(2h)$. Analogously, the expressions for the corresponding lumped circuit parameters of this structure can be readily obtained by replacing $m\pi$ in Eqs. (5) – (8) with $\xi_m$ .Substituting these expressions into Eq. (2) shows that Eq. (9) remains applicable for calculating the Q-factors of arbitrary-order MP resonances. Nevertheless, the substrate-induced asymmetry leads to unequal current loops and nonuniform loss distribution, requiring a correction factor $c_1$.

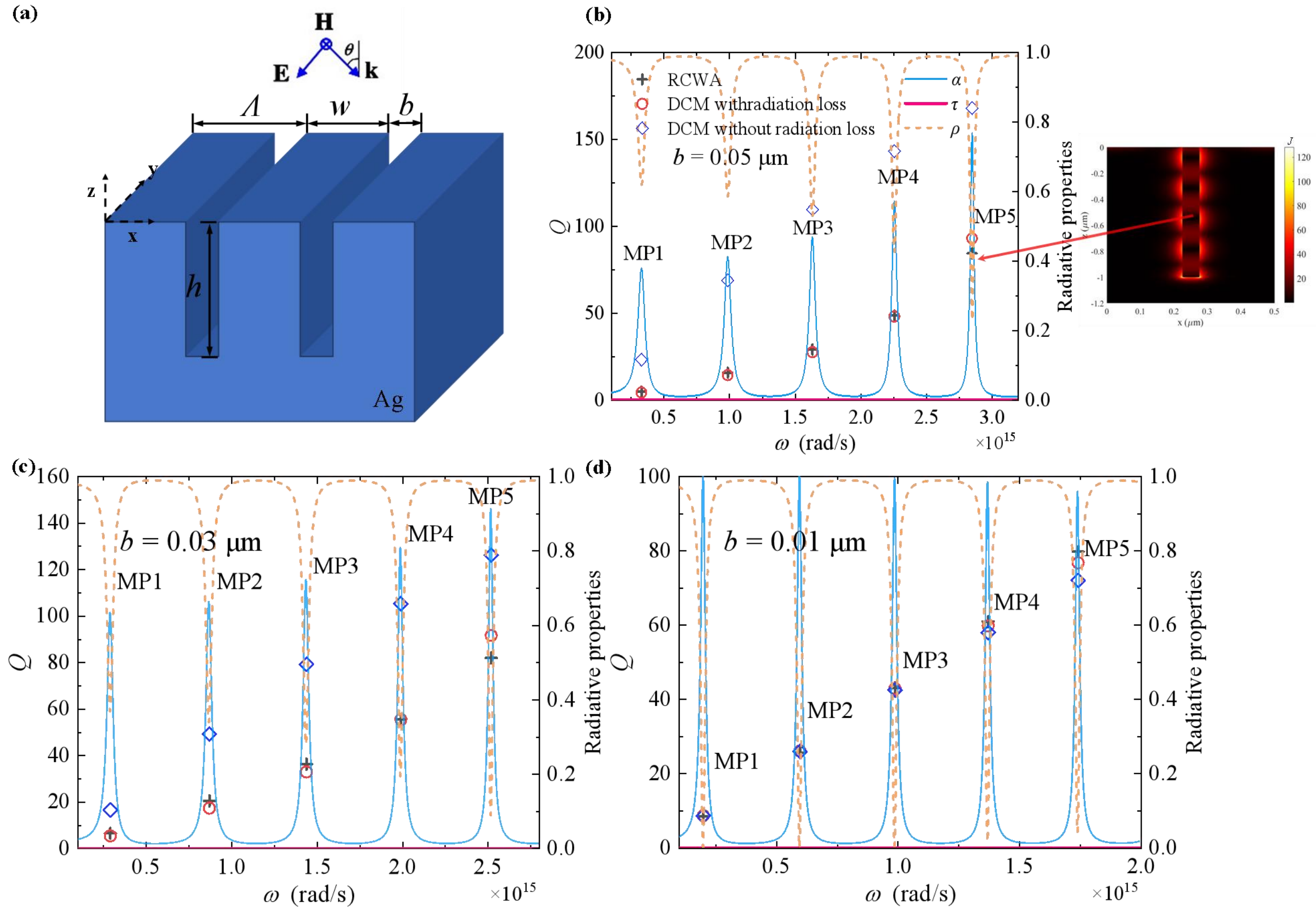

**Fig. 4** Validation of the DCM predictions for Q-factors of multi-order MPs in substrate-supported vertical MIM structures. (a) Schematic of the substrate-supported vertical MIM structure; (b) Results at Λ=0.5μm, $h$=1μm, and $b$=0.05μm with $c_1$=2.6; (c) Results at $b$=0.03μm with $c_1$ = 3.5; (d) Results at b=0.01μm with $c_1$ = −1.4. All results are obtained under normal incidence.

The validation results are presented in **Fig. 4**. With Λ = 0.5 and $h$ = 1μm fixed under normal incidence, **Fig. 4**(b) presents the prediction results for $b$ = 0.05μm, where $c_1$ = 2.6. The MRE decreases from 254.18% to 7.33% when radiation loss is included. **Fig. 4**(c) shows the corresponding results for $b$ = 0.03μm, where $c_1$ = 3.5. When the slit width is further reduced to $b$ = 0.01μm, the correction coefficient is a negative value, $c_1$ = −1.4. For all three structures considered, the Q-factor prediction behavior at small slit widths differs significantly from that at larger slit widths. For the substrate-free vertical MIM structure, the value of $c_1$ deviates from that corresponding to a uniform distribution of loss energy when the slit width becomes small. For both the substrate-supported vertical and horizontal MIM structures, $c_1$ becomes negative at small slit widths. At this time, when radiative losses are neglected, the predicted Q-factors are smaller than those obtained from RCWA, indicating that the interaction between radiative and Ohmic losses can no longer be described as a simple linear superposition. Instead, a strong coupling effect emerges between these two loss channels.

In summary, a DCM incorporating radiative losses was developed to predict the Q-factors of arbitrary-order MPs in various MIM structures. By introducing a radiation resistance and establishing a mapping between distributed and lumped parameters, a unified analytical expression for Q-factors was derived. The model shows excellent agreement with RCWA calculations over a broad range of geometries. The correction coefficient $c_1$ effectively accounts for nonuniform loss distributions among multiple current loops. Furthermore, identical Q-factors are obtained from the absorptance, reflectance, and transmittance spectra, confirming their common physical origin from the same MP resonance. Although further developments are required to describe strong radiative–Ohmic coupling and MP–SPP hybridization, the proposed model provides a simple yet powerful framework for designing high-performance metamaterial emitters and absorbers with tailored spectral selectivity. More importantly, it offers deeper physical insight into the loss mechanisms governing MP-based micro/nanophotonic systems.

## Acknowledgment

JMZ thanks the support by the National Natural Science Foundation of China (No. 51976045) and the Young Scientist Studio Foundation of Harbin Institute of Technology.

## Data availability statement

The data that support the findings of this study are available from the corresponding author upon reasonable request.

## References

[1] D. G. Baranov, Y. Xiao and I. A. Nechepurenko et al., "Nanophotonic Engineering of Far-Field Thermal Emitters," Nat. Mater. **18**, 920-930 (2019).

[2] Y. Matsuno and A. Sakurai, "Electromagnetic resonances of wavelength-selective solar absorbers with film-coupled fishnet gratings," Opt. Commun. **385**, 118-123 (2017).

[3] W. W. Zhang, H. Qi and Y. M. Yin et al., "Tailoring radiative properties of a complex trapezoidal grating solar absorber by coupling between SPP and multi-order MP for solar energy harvesting," Opt. Commun. **479**, 126416 (2021).

[4] J. Su, D. Liu and L. Sun et al., "Plasmonic Nanostructures for Broadband Solar Absorption Based on Synergistic Effect of Multiple Absorption Mechanisms," Nanomaterials **12** (24), 4456 (2022).

[5] C. C. Chang, W. J. M. Kort-Kamp and J. Nogan et al., "High-Temperature Refractory Metasurfaces for Solar Thermophotovoltaic Energy Harvesting," Nano Lett. **18** (12), 7665-7673 (2018).

[6] B. Zhao, L. Wang and Y. Shuai et al., "Thermophotovoltaic emitters based on a two-dimensional grating/thin-film nanostructure," Int. J. Heat Mass Transf. **67**, 637-645 (2013).

[7] M. A. Abbas, J. Kim and A. S. Rana et al., "Nanostructured chromium-based broadband absorbers and emitters to realize thermally stable solar thermophotovoltaic systems," Nanoscale **14** (17), 6425-6436 (2022).

[8] S. Zeng, S. Pian and M. Su et al., "Hierarchical-morphology metafabric for scalable passive daytime radiative cooling," Science **373** (6555), 692-696 (2021).

[9] D. Li, X. Liu and W. Li et al., "Scalable and hierarchically designed polymer film as a selective thermal emitter for high-performance all-day radiative cooling," Nat. Nanotechnol. **16** (2), 153-158 (2021).
[10] L. Long, S. Taylor and L. Wang, "Enhanced Infrared Emission by Thermally Switching the Excitation of Magnetic Polariton with Scalable Microstructured VO2 Metasurfaces," ACS Photonics **7** (8), 2219-2227 (2020).
[11] Y. Liu, J. Song and W. Zhao et al., "Dynamic thermal camouflage via a liquid-crystal-based radiative metasurface," Nanophotonics (Berlin, Germany) **9** (4), 855-863 (2020).
[12] H. Zhu, Q. Li and C. Tao et al., "Multispectral camouflage for infrared, visible, lasers and microwave with radiative cooling," Nat. Commun. **12** (1), 1805 (2021).
[13] B. Qin, Y. Zhu and Y. Zhou et al., "Whole-infrared-band camouflage with dual-band radiative heat dissipation," Light: Science & Applications **12** (1), 246 (2023).
[14] R. Feng, J. Qiu and L. Liu et al., "Parallel LC circuit model for multi-band absorption and preliminary design of radiative cooling," Opt. Express **22** (S7), A1713 (2014).
[15] B. Zhao and Z. M. Zhang, "Study of magnetic polaritons in deep gratings for thermal emission control," Journal of Quantitative Spectroscopy and Radiative Transfer **135**, 81-89 (2014).
[16] L. P. Wang and Z. M. Zhang, "Resonance transmission or absorption in deep gratings explained by magnetic polaritons," Appl. Phys. Lett. **95** (11), 111904 (2009).
[17] H. Li and J. Zhao, "Predicting Multi-Order Magnetic Polariton Resonances for Radiative Properties Tailoring by Distributed Circuit Model," Int. J. Heat Mass Transf. **256** (2026).
[18] L. P. Wang and Z. M. Zhang, "Effect of magnetic polaritons on the radiative properties of double-layer nanoslit arrays," Journal of the Optical Society of America B **27**, 2595-2604 (2010).
[19] L. P. Wang and Z. M. Zhang, "Phonon-mediated magnetic polaritons in the infrared region," Opt. Express **19 Suppl 2**, A126-A135 (2011).
[20] T. Søndergaard, J. Jung and S. I. Bozhevolnyi et al., "Theoretical analysis of gold nano-strip gap plasmon resonators," New J. Phys. **10** (10), 105008 (2008).
[21] J. Park, J. H. Kang and X. Liu et al., "Electrically Tunable Epsilon-Near-Zero (ENZ) Metafilm Absorbers," Sci. Rep. **5** (1), 15754 (2015).
[22] A. Pors, O. Albrektsen and I. P. Radko et al., "Gap plasmon-based metasurfaces for total control of reflected light," Sci. Rep. **3** (1), 2155 (2013).
[23] J. Kim, A. Dutta and B. Memarzadeh et al., "Zinc Oxide Based Plasmonic Multilayer Resonator: Localized and Gap Surface Plasmon in the Infrared," ACS Photonics **2** (8), 1224-1230 (2015).
[24] F. Ding, Y. Yang and R. A. Deshpande et al., "A review of gap-surface plasmon metasurfaces: fundamentals and applications," Nanophotonics **7** (6), 1129-1156 (2018).
[25] Y. Gong, Y. Guo and Z. Liu et al., "Target-oriented spectral emissivity design: Mechanism and prediction of magnetic polaritons based on transmission line theory," Int. J. Heat Mass Transf. **244**, 126904 (2025).
[26] A. Sakurai, B. Zhao and Z. M. Zhang, "Resonant frequency and bandwidth of metamaterial emitters and absorbers predicted by an RLC circuit model," Journal of Quantitative Spectroscopy and Radiative Transfer **149**, 33-40 (2014).
[27] M. Dareini, S. R. Ghorbani and H. Arabi et al., "Application of Circuit Model for Gap-Plasmon Nanodisk Resonators," Photonics Nanostruct. **60** (2024).
[28] J. C. Zhao, Y. M. Dang and H. Y. Hou et al., "Design and Validation of RLC Equivalent Circuit Model Based on Long-Wave Infrared Metamaterial Absorber," J. Infrared Millim. Waves **44** (2025).
[29] X. Xu, Y. Li and B. Wang et al., "Prediction of Multiple Resonance Characteristics by an Extended Resistor-Inductor-capacitor Circuit Model for Plasmonic Metamaterials Absorbers in Infrared.," Opt. Lett. **40** (2015).
[30] L. Long, Y. Yang and H. Ye et al., "Optical absorption enhancement in monolayer MoS2 using multi-order magnetic polaritons," Journal of Quantitative Spectroscopy and Radiative Transfer **200**, 198-205 (2017).
[31] Y. Guo, B. Xiong and Y. Shuai, "Predicting Multi-Order Magnetic Polaritons Resonance in SiC Slit Arrays by Mutual Inductor–Inductor–Capacitor Circuit Model," J. Heat Transf.-Trans. ASME **142** (7), 72801 (2020).

[32] Y. Guo, S. Zhou and B. Xiong et al., "Mechanism and prediction of multi-mode magnetic polaritons by MCLC circuit model in complex micro/nanostructures," Journal of Quantitative Spectroscopy and Radiative Transfer **269**, 107700 (2021).
[33] E. D. Palik and G. Ghosh, (Academic Press, San Diego, CA, 1999).
[34] Y. B. Chen and Z. M. Zhang, "Design of tungsten complex gratings for thermophotovoltaic radiators," Opt. Commun. **269** (2), 411-417 (2007).
[35] Y. Guo, Y. Shuai and J. Zhao, "Tailoring Radiative Properties with Magnetic Polaritons in Deep Gratings and Slit Arrays Based on Structural Transformation," J. Quant. Spectrosc. Radiat. Transf. **242**, 106788 (2019).
[36] Z. M. Zhang, *Nano/microscaleheattransfer*. (Springer, Cham, 2020).